# Damping Reveals Hidden Dimensions in Elastic Metastructures Through Induced Transparency

Yanghao Fang,[1, *] William Tuxbury,[2, *] Abhishek Gupta,[3] Tsampikos Kottos,[2, †] Ramathasan Thevamaran[1, 3, ‡]

[1]*Department of Materials Science and Engineering, University of Wisconsin-Madison, Madison, Wisconsin 53706, USA*
[2]*Wave Transport in Complex Systems Lab, Department of Physics, Wesleyan University, Middletown, Connecticut 06459, USA*
[3]*Department of Mechanical Engineering, University of Wisconsin-Madison, Madison, Wisconsin 53706, USA*

[*] These authors contributed equally to this work
[†] tkottos@wesleyan.edu
[‡] thevamaran@wisc.edu

Damping typically results in attenuation of vibrations and elastic wave propagation in mechanical systems. Contrary to this conventional understanding, we demonstrate experimentally and explain theoretically the revival of an elastic wave transmitted through a periodic metastructure when a weak non-Hermitian defect (damping mechanism) induces violation of time-reversal symmetry. Damping alters the nature of the system's resonant modes, instigating interference in the scattering field. This leads to transmission revival, revealing the presence of hidden modes which are otherwise masked by the symmetry. Our findings offer an innovative approach for designing dissipation driven switches and controllers and non-destructive structural health monitoring systems.

## I. INTRODUCTION

Defects in the geometry or mass of a periodic system have commonly been used to elicit distinct elastic/acoustic wave phenomena not present in perfectly periodic configurations [1] such as one-way wave propagation [2], bifurcation-based mechanical rectification [3], mechanical energy localization [4], non-linear acoustic diodes [5], bounded modes [6,7], and topologically protected states [8,9]. While these theories predominantly focus on conservative (Hermitian) defects which violate the translational invariance, the introduction of localized energy amplification and/or attenuation defect mechanisms (non-Hermitian) [10,11] violates also the time-reversal symmetry, leading to unexpected phenomena such as anomalous actuation force (emissivity) enhancement [12], hypersensitive sensing [13–18], shadow-free sensing [19], unidirectional invisibility [20,21] and more. For recent reviews on the implementation of symmetries (and their violation) in non-Hermitian structures in a variety of frameworks, including pressure acoustics, see references [22,23].

Here, we demonstrate experimentally and explain theoretically that integrating a weak non-Hermitian defect—characterized by energy dissipation or damping—into an otherwise Hermitian metastructure can lead to a counter-intuitive phenomenon: the revival of the longitudinal wave transmission when the system is excited from its plane of parity symmetry as illustrated in Fig. 1(a). This observation challenges the conventional understanding that damping is an attenuation mechanism and extends the utility of damping beyond its common use in wave attenuation or vibration mitigation. In fact, similar unexpected phenomena have been observed in the photonics framework where the gain reduction (by added loss) leads to lasing action [10,24,25].

In our mechanical paradigm, impulse excitation experiments (Fig. 1(b)) guided by finite element (FE) analysis and coupled-mode-theory (CMT) modeling reveal a loss-induced resurgence of transmission at frequencies corresponding to anti-symmetric eigenmodes of the metastructure. While the spatial parity symmetry violation is observable from the destruction of the nodal point, our theoretical model points to a profound insight: the irreversibility introduced by the damping violates the time-reversal symmetry of the metastructure, introducing a phase-lag to the motion of the individual components of the metastructure. The alteration of the eigenmode nature (from real to complex valued components), reinstates the contribution of a hidden mode in the transmission leading to its revival that is anomalously enhanced in the weak damping regime. Specifically, the CMT analysis predicts a linear growth of transmittance for weak damping strengths compared to, for example, the enhancement due to purely geometric (non-dissipative) defects which demonstrate a (weaker) quadratic growth of transmittance in the limit of small perturbations.

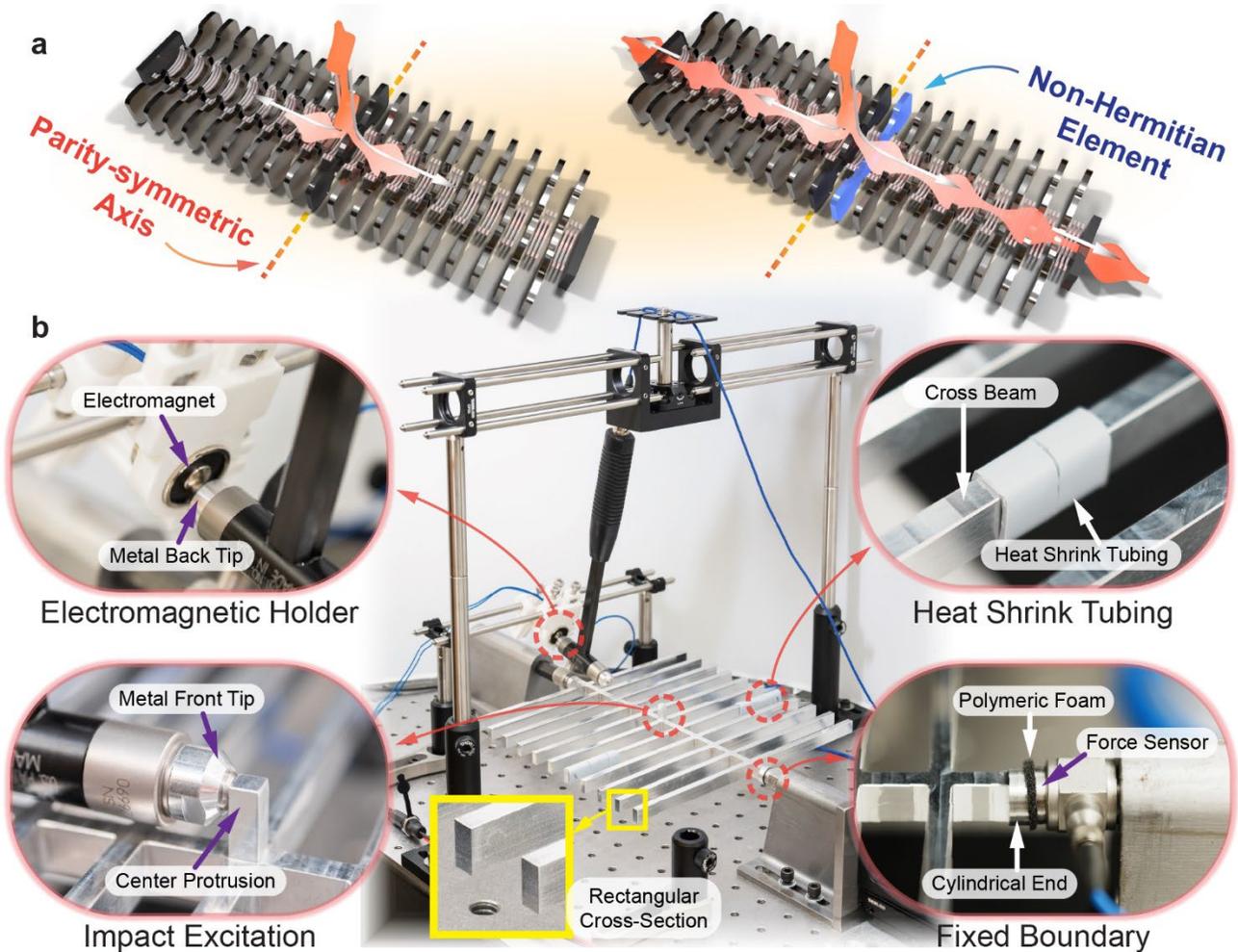

FIG. 1. Illustrations and experimental setup pictures of transmission revival in a periodic metastructure with a non-Hermitian defect. (a) Transmission signal variation before (left) and after (right) introducing a non-Hermitian defect when the system is excited from its center. (b) A custom-built impulse experiment apparatus showing an electromagnetic holding/releasing mechanism for the repeatable impulse excitation by a hammer (top left inset), the impulse excitation by the instrumented hammer at the center of metastructure (bottom left inset), the heat shrink tubing on the cross beam used for damping (top right inset), and the metastructure boundaries supported by dynamic force sensors on either side with a polymeric foam placed between them to achieve uniform contact for transmission measurements (bottom right inset). The bottom middle inset shows the rectangular cross-section of the beam.

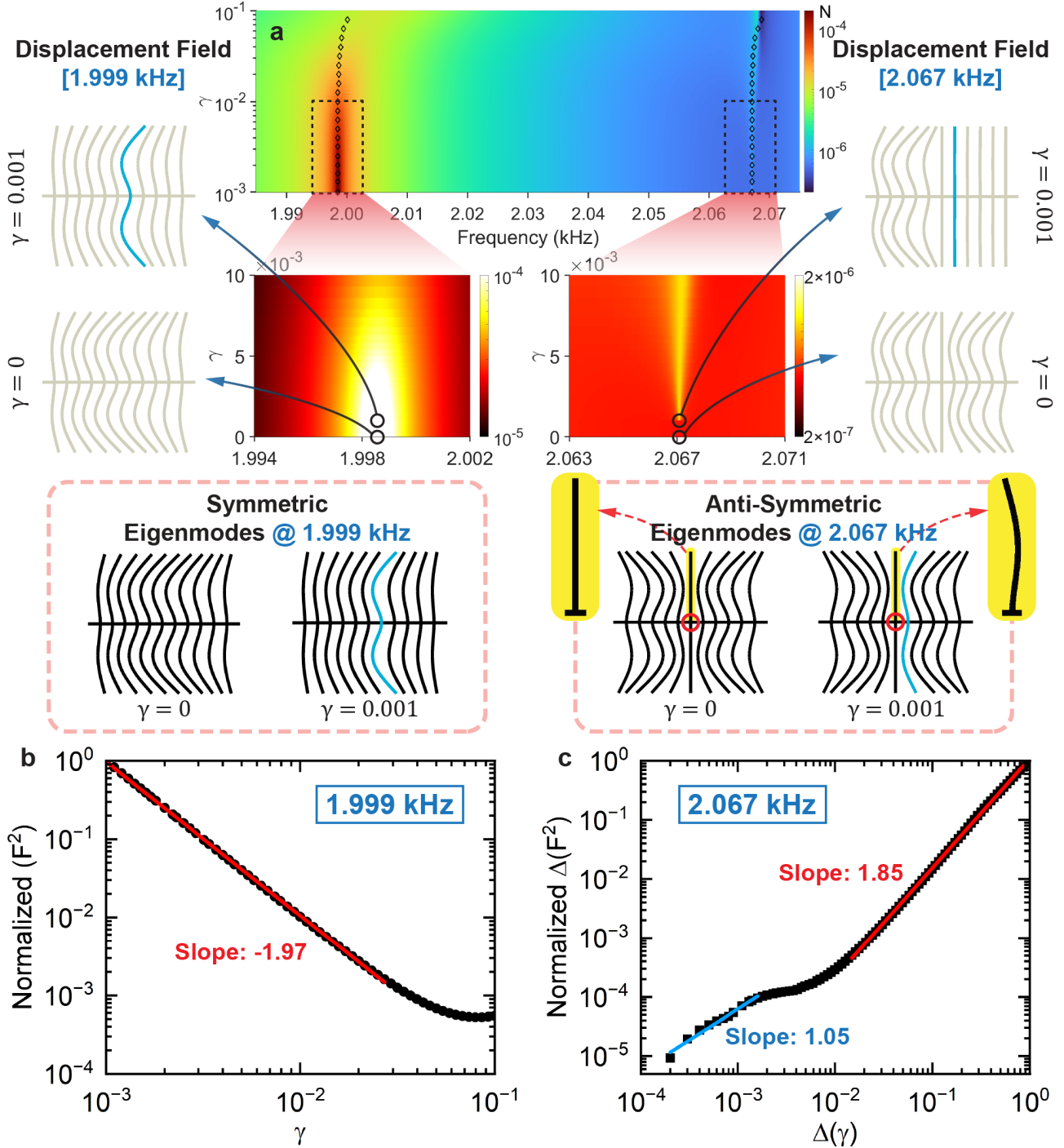

FIG. 2. Transmitted signals under center excitation of the metastructure. (a) Color density plots (using FE analysis) of transmitted force overlaid by eigenvalue calculations (represented as black hollow diamonds) between 1.985 kHz and 2.085 kHz as the damping intensity $\gamma$ varies between 0.001 and 0.1. The insets on the left and right sides illustrate the displacement fields (non-Hermitian components are in blue) for two modes located at 1.999 kHz and 2.067 kHz when $\gamma = 0$ (Hermitian) and $\gamma = 0.001$ (non-Hermitian), respectively. The corresponding eigenmodes are shown below enclosed in red dashed boxes. For the eigenmodes at 2.067 kHz, the red circles indicate the locations of the center nodes, and the yellow regions depict the slight deformation difference of the center beam as $\gamma$ increases from 0 to 0.001. (b) Square of the normalized force intensity (normalized $F^2$, which is proportional to the transmitted power) versus damping intensity $\gamma$ for the mode at 1.999 kHz. A fitting slope of -1.97 represents the decay ratio. (c) Variation in the square of the normalized output force for the mode at 2.067 kHz

(denoted as $\Delta(F^2) \equiv F^2 - F^2_{\gamma=0.001}$) plotted against the increase in damping ($\Delta(\gamma) \equiv \gamma - 0.0009$). The slopes within the weak damping and strong damping regions are $1.05 \pm 0.05$ and $1.854 \pm 0.005$, respectively.

## II. RESULTS AND DISCUSSIONS

### 1. Finite element modeling

We use steady-state dynamic analyses in the commercial FE software *ABAQUS Simulia* for modeling a periodic metastructure featuring a non-Hermitian defect. This metastructure comprises $2N + 1$ cross beams, with $N$ cross beams on each side and one in the center. In FE modeling, we use an $N_5$ (i.e., $N = 5$) periodic metastructure which has 5 elastic cross beams on each side, and the 1st cross beam on the right side counting outwards from the center (i.e., the 7th cross beam counting from the left side) has a damping mechanism that introduces the non-Hermiticity to the system. Each cross-beam section, extending from the tip to its junction with the horizontal beam, is 110 mm in length, with adjacent sections separated by a 20 mm center-to-center gap. The horizontal beam has 20 mm ledges on both ends. The whole metastructure is defined as aluminum (with a Young's modulus of 68.9 $GPa$, a density of 2700 $kg\ m^{-3}$, and a Poisson's ratio of 0.33) and contains a uniform rectangular cross-section (5 mm wide and 10 mm thick as shown in the yellow inset in Fig. 1(b)). We use 3-node quadratic Timoshenko beam elements for the mesh, and both the cross-beam sections and the interval segment of the horizontal beam comprises twenty elements. The loss mechanism is modeled by structural damping rate $\gamma$.

To study how the strength of non-Hermiticity (damping strength $\gamma$) affects the transport, we calculate the reaction force signal under center excitation (parity symmetry plane of the metastructure) as we increase the damping intensity. To further understand the effects of the excitation conditions, we have also performed a study under edge excitation of which the results can be found in Fig. 4 in Appendix A. The center excitation is operated when the metastructure is excited harmonically with a prescribed 1 $\mu N$ longitudinal force at the metastructure's center while the ends are kept fixed, the transmission signal is collected from the fixed end located on the side without the damping component; the edge excitation is operated when one edge of the metastructure is excited harmonically with prescribed 1 $\mu m$ longitudinal displacement while the other end remained fixed (i.e., zero displacement and rotation), the transmission signal is collected from the fixed end.

Color density plots of the transmission response corresponding to the center excitation can be seen in Fig. 2(a), together with the displacement fields and selective eigenmodes. The reaction force magnitude was calculated from the fixed left end where there is no damping element, and the eigenmodes of the metastructure are calculated under the boundary conditions where both ends are fixed. The two eigenvalues that exist between 1.985 and 2.085 kHz are represented by the black diamonds shown atop the density plot and their trajectories overlap with the resonance peak locations found in the FE model. The insets on either side show the displacement fields of the two modes in the spectrum for two values of $\gamma = 0$ and $\gamma = 0.001$. The left column shows the displacement fields of a (typical) mode at 1.999 kHz, which demonstrates the same features with its corresponding eigenmodes. For $\gamma = 0$ the eigenmode is symmetric since its corresponding elements on both parity-symmetric (P-symmetric) halves of the metastructure move in the same direction simultaneously, thus respecting the symmetry of the metastructure (as expected). This similarity between the displacement field and the eigenmode persists also for $\gamma = 0.001$. In contrast, the displacement fields corresponding to the mode at 2.067 kHz are different from the associated eigenmode for both $\gamma = 0$ and $\gamma = 0.001$ as evident in the right insets of Fig. 2(a). Specifically, for $\gamma = 0$ the eigenmode is anti-symmetric (respecting the parity-symmetry of the metastructure) with its two P-symmetric halves moving in the opposite directions, resulting in a nodal point at the center (highlighted with a red circle). Its displacement field, however, is symmetric due to the center excitation disrupting the nodal point. When a small amount of damping is introduced ($\gamma = 0.001$) the eigenmode maintains its overall parity behavior (strictly speaking, a small imaginary part of the components is acquired, and the center beam starts to deform slightly as shown in the yellow regions near the anti-symmetric eigenmodes in Fig. 2(a)) while the corresponding displacement field experiences an abrupt parity violation originating from a superposition of symmetric and anti-symmetric eigenmodes.

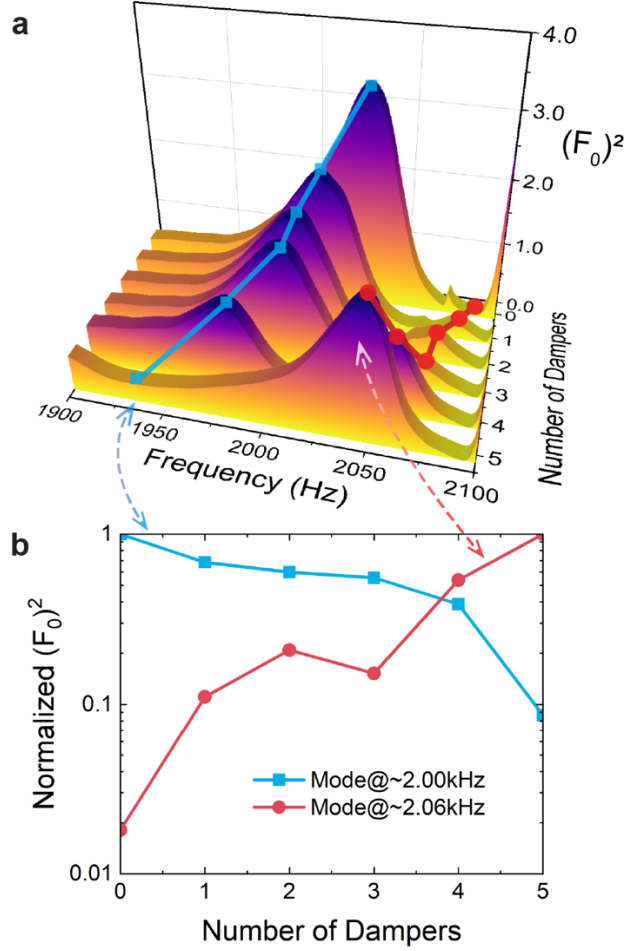

FIG. 3. Transmission revival in experiments. (a) The square of the transmission response between 1.9 and 2.1 kHz of the metastructure versus number of dampers. The amplitude $(F_0)^2 = (|F_s| \cdot |F_h|^{-1})^2$ is proportional to the transmitted power, with $F_s$ and $F_h$ denoting signal amplitudes from the force sensor and hammer in their Fourier representations, respectively. (b) Variation of the amplitudes $(F_0)^2$ corresponding to the two distinct modes near 2.00 kHz and 2.06 kHz (normalized by their corresponding local maxima) as functions of number of dampers.

For a better understanding of the variation of the transmitted power—which is proportional to the square of the transmitted force, $F^2$—with respect to damping intensity $\gamma$, Fig. 2(b) presents the mode at 1.999 kHz. The fitting slope of -1.97 implies that $F^2 \propto \gamma^{-1.97}$, indicating a conventional transmitted power decay due to damping. Fig. 2(c) shows how the transmitted power of the 2.067 kHz mode varies with increasing damping. For weak damping, the variation behaves like $\Delta(\gamma)^\alpha$, where the value of $\alpha = 1.05 \pm 0.05$ was evaluated from a least-square fitting after extracting the asymptotic ($\gamma \to 0$) value of the transmittance from the *ABAQUS Simulia* simulations. For stronger damping, the least-square fit of the *ABAQUS Simulia* data of the square of the transmitted force indicates a power law variation which behaves like $\Delta(\gamma)^\alpha$, where $\alpha = 1.854 \pm 0.005$. These results are in qualitative agreement with the CMT prediction of transmittance for the frequency associated with the anti-symmetric mode. In fact, the CMT predicts a linear variation for small $\gamma$ values. As loss increases, the resonant detuning (real part) of the defect mode frequency dominates the behavior of the transmittance. In the extreme limit where the defect resonant variation is completely dominated by its real part, CMT predicts a quadratic behavior of the transmittance with respect to variations of the defect magnitude, in agreement with the *ABAQUS Simulia* simulations (see Sec. CMT Model below and Appendix B). It is worth noting that such behavior occurs only at the frequencies corresponding to anti-symmetric modes, and the symmetric modes are attenuated similar to the behavior of 1.999 kHz mode. In P-symmetric metastructures excited by longitudinal waves, eigenmodes alternate between symmetric and anti-symmetric profiles. Consequently, this characteristic of transmission revival persists across different spectral bands, system sizes, and defect locations excluding the nodal point corresponding to an anti-symmetric mode (see Fig. 5 and Fig. 6 in Appendix A). If the defect is located at the nodal point corresponding to a particular anti-symmetric mode, there will be no transmission revival at that mode's frequency. Any other defect location can lead to transmission revival.

## 2. Impulse experiments

The metastructure is fabricated from monolithic aluminum 6061-T6 using CNC machining under the basic geometry in FE model. As shown in Fig. 1(b), our sample was placed between two dynamic force sensors (PCB 208C01). Considering the shape of the force sensors, we designed the tips of the main beam's ends with two flat cylinders, each 12.7 mm in diameter, while keeping the total length consistent. We applied a precompression to prevent loss of contact with the force sensors and sliding, and introduced protrusions at the center of the main beam for center excitation of the metastructure with an instrumented impact hammer (PCB 086C01). To accurately control the hammer impulse with reproducibility, we devised a rotational mechanical hammer suspension mechanism coupled with an electromagnetic hammer hold/release system as shown in the insets of Fig. 1(b). We introduce damping to both sections of the 7$^{th}$ cross beam of the metastructure counting from the left side by affixing shrunken heat shrink tubing to the tips. Each tubing segment weighs 0.422 grams (±1%), accounting for roughly 2.8% of the aluminum cross beam section's weight of 14.85 grams.

We measure time-domain signals from both the instrumented hammer and the dynamic force sensors during the impact. After denoising by reserving only the pulse signal from the hammer and removing signals prior to the impact from the force sensor, we use Fast Fourier Transform to transform these signals to the frequency domain. The hammer impulse excites frequencies up to approximately 3 kHz. We subsequently normalize the force sensors' signal where the normalized force $F_0 = |F_s| \cdot |F_h|^{-1}$, with $F_s$ and $F_h$ denoting signal amplitudes from the force sensor and hammer in their Fourier representations, respectively.

Fig. 3 shows that the transmitted power of the mode near 2.00 kHz diminishes and shifts slightly to a lower frequency as the damping is increased, indicating a typical energy attenuation. In contrast, the mode at 2.06 kHz appears spontaneously when the damping is introduced, and its transmitted power increases rapidly as the damping is increased further— signifying a transmission revival phenomenon. Specifically, the amplitude of the transmitted power increases by two orders of magnitude from the "0 dampers" (i.e., no dampers) to "5 dampers" (i.e., damping introduced by 5 heat-shrink tubing dampers on both ends of the cross beam). It is noteworthy that introducing damping using heat-shrink tubing will inevitably add extra mass (~2.8% per tubing) to the defect beam, which causes some shift in the mode frequency, but we show later in our CMT analysis that the effects due to such geometric defect is small compared to the introduced damping.

## 3. CMT model

To describe the theoretical underpinnings of the computational and experimental results presented in the previous sections, we have developed a simplified model based on CMT [26], which allows us to theoretically explain the observed strong transmission revival for weak dissipations. The fundamental modes associated with the cantilever free vibration of the isolated cross beams can be modeled as a one-dimensional array of $2N + 1$ isolated resonances with frequency $\omega_0$ and nearest neighbor coupling $\kappa$. Meanwhile, non-Hermiticity introduced by damping on the $N + l + 1$ element can be incorporated by the imaginary term $-i\gamma$. Therefore, the Hamiltonian operator of the closed system is expressed in the form, $\hat{H} = \sum_{n=1}^{2N+1} \omega_0 |n\rangle\langle n| + \sum_{n=1}^{2N} \kappa |n\rangle\langle n+1| + c.c. - i\gamma |N + l + 1\rangle\langle N + l + 1|$, where $|n\rangle$ indicates the isolated mode of the $n^{th}$ cross beam. The system can then be turned to a scattering setup by attaching transmission lines (TLs) to the respective isolated resonances, which are accounted for by incorporating an imaginary self-energy correction to those diagonal matrix elements, leading to the effective

Hamiltonian, $\widehat{H}_{eff} = \widehat{H} - \frac{i}{2}\widehat{D}\widehat{D}^T$. Here $\widehat{D} = \sqrt{2\gamma_e}(|1\rangle\langle e_1| + |N+1\rangle\langle e_2| + |2N+1\rangle\langle e_3|)$ is referred to as the coupling matrix, where $|e_m\rangle$ indicates the flux-normalized propagating modes in the $m^{th}$ TL and $\gamma_e$ encodes the coupling coefficient to those isolated resonances.

Using a Green's function approach, we are able to acquire analytic expressions for the transmittance to the leftmost TL, due to a center-excitation at frequency $\omega = \omega_0$ with respect to the non-Hermitian defect perturbation, where tildes indicate the quantities are normalized by the nearest neighbor coupling $\kappa$,

$$T(\tilde{\gamma}) = \left[\frac{2\tilde{\gamma}_e(\tilde{\gamma}_e+\tilde{\gamma})}{(\tilde{\gamma}_e+\tilde{\gamma})(\tilde{\gamma}_e^2+1)+\tilde{\gamma}_e}\right]^2 \quad (1)$$

Inspection of the formula leads to the immediate conclusion that *the variation in the transmittance is first order in the damping parameter*, $T(\tilde{\gamma}) \approx T(0) + \frac{8\tilde{\gamma}_e}{(2+\tilde{\gamma}_e)^3}\tilde{\gamma}$. In the case of zero damping perturbation, there is a background transmittance $T(0) \neq 0$ associated with the contribution of nearby (symmetric) resonances of the composite metastructure, referred to as the super-modes, at which point the hidden mode is undetectable. In the derivation, we have assumed $N$ and $l$ are both odd, so the defect resides on an anti-node of the unperturbed anti-symmetric super-mode. Otherwise, the expression is independent of the system size and the defect's position.

To highlight the importance of the (weak) dissipative defect in the observed (strong) transmittance revival, we contrast the above results with a Hermitian (geometric) non-dissipative defect $\Delta \in \mathbb{R}$. In this case, the transmittance takes the form, $T(\widetilde{\Delta}) = \frac{4\tilde{\gamma}_e^2(\tilde{\gamma}_e^2+\widetilde{\Delta}^2)}{(\tilde{\gamma}_e^2+\widetilde{\Delta}^2)(\tilde{\gamma}_e^2+1)^2+2\tilde{\gamma}_e^2(\tilde{\gamma}_e^2+1)+\tilde{\gamma}_e^2}$, also following the same tilde notational convention as in the non-Hermitian case. The variation in the transmittance in the case of a Hermitian defect turns out to be a second order effect of the defect strength $\widetilde{\Delta}$, i.e., $T(\widetilde{\Delta}) \approx T(0) + \frac{4(3+2\tilde{\gamma}_e^2)}{(2+\tilde{\gamma}_e)^4}\widetilde{\Delta}^2$ which for weak perturbations is much smaller than the previous results for the dissipative defects. Such comparative analysis indicates that time-reversal symmetry violation plays a significant role in uncovering the hidden mode via its contribution to the transmittance.

In Appendix B, we analyze in detail the effect that nodal points of anti-symmetric eigenmodes have on the transmittance and relate the phenomena to a single-sided resonance [27,28]. Additionally, we use first order non-degenerate perturbation theory to study the impact of a non-Hermitian defect on the eigenfrequency and nodal points of the unperturbed anti-symmetric eigenmode. The calculation demonstrates that, in addition to the P-symmetry violation which fundamentally disrupts the nodal point, broken time-reversal symmetry introduces a phase change to the components of the eigenmodes, resulting in interference between the super-modes comprising the scattering field and leading to an enhanced (over Hermitian perturbations) transmittance revival. An important benefit of these dissipative perturbations is the fact that they do not distort the resonant frequency while only increasing the resonance linewidth. The theoretical predictions provided by the CMT model underscore the utility of violating hidden (time-reversal) symmetries in the design and application of wave mechanical devices based on symmetry management techniques.

## III. CONCLUSIONS

We have experimentally, computationally, and theoretically studied the transmission properties of an elastodynamic metastructure with an embedded non-Hermitian (damping) defect. We observed an emergence of the transmittance at frequencies associated with anti-symmetric eigenmodes of the underlying Hermitian metastructure upon tuning the damping strength of the defect revealing those hidden dimensions of the system. This phenomenon is independent of the system's size, the spectral band, and the location of defect as long as the defect does not lie at a nodal point. Our CMT model demonstrates that integration of a non-Hermitian defect enables otherwise obstructed energy transport by disrupting the central nodal point of anti-symmetric eigenmodes without shifting their frequency. Comparison to a purely geometric defect shows that violation of the hidden time-reversal symmetry by the damping element dramatically intensifies the resurgence for weak perturbations—a linear enhancement in contrast to quadratic enhancement achievable with geometric defects. These findings show an intriguing counter-intuitive approach for inducing mechanical wave transparency by introducing energy loss mechanisms to an otherwise Hermitian system. Our results underscore the potential of non-Hermitian physics in reshaping our understanding and ability to manipulate wave propagation in elastodynamic systems. Our proof-of-principle experiments have technological ramifications for applications in non-destructive structural health monitoring, and realization of dissipation driven switches and controllers, especially when dissipative nonlinearities self-induce symmetry violations.

## ACKNOWLEDGEMENTS

Y.F., A.G., and R.T. acknowledge the financial support from the UW-Madison MRSEC program (Grant No. 1720415), the DCSD program of the National Science Foundation (Grant No. NSF-CMMI-1925530), and the Army Research Office (Grant No. W911NF2010160). We thank Professor Melih Eriten for the instrumented impact hammer.

W.T. and T.K. acknowledge the financial support from the DCSD program of the National Science Foundation (Grant No. NSF-CMMI-1925543) and the Simons Foundation for Collaboration in MPS No. 733698.

## APPENDIX A: FINITE ELEMENT ANALYSIS

This appendix provides results from the FE analysis. Figure 4 presents the transmitted signal from edge excitation of the same metastructure featured in the manuscript. The displacement fields at 1.999 kHz and 2.067 kHz correspond closely to the eigenmodes displayed in Fig. 2(a).

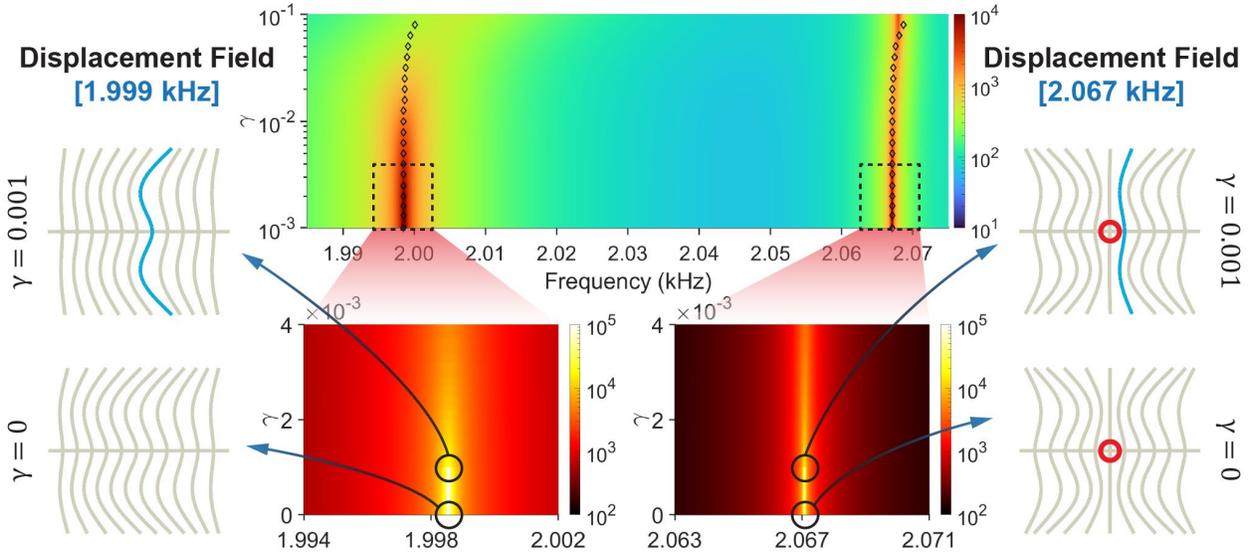

FIG. 4. Transmitted signal under edge excitation of the metastructure. Color density plots (using FE analysis, unit is in Newtons) overlaid by eigenvalue calculations (represented as black hollow diamonds) between 1.985 kHz and 2.085 kHz as the damping intensity $\gamma$ varies between 0.001 and 0.1. The insets on the left and right sides illustrate the displacement fields (non-Hermitian components are in blue) for modes at 1.999 kHz and 2.067 kHz at $\gamma = 0$ (Hermitian) and $\gamma = 0.001$ (non-Hermitian), respectively. They are consistent with their eigenmodes due to their identical geometrical configuration.

Figures 5 and 6 show the transmitted signal from center excitation of a slightly different metastructure. These results demonstrate that the phenomenon of transmission revival can be observed across different spectral bands, system sizes, and defect locations.

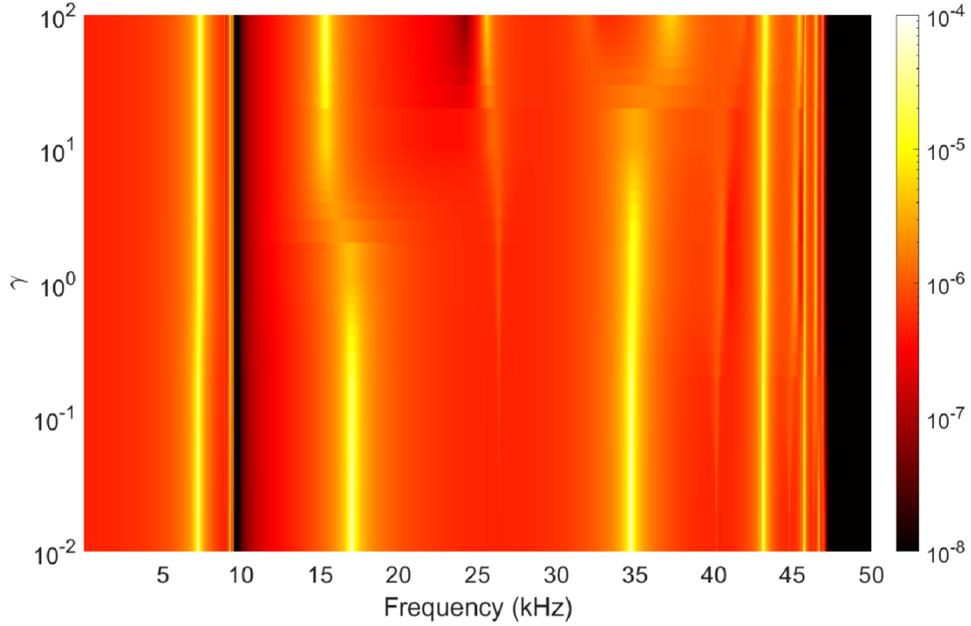

FIG. 5. Density plot of the transmitted signal for an $N_5$ metastructure (slightly different from the one used in the manuscript) under center excitation with a non-Hermitian defect (where damping intensity $\gamma$ is added) on its $1^{st}$ cross beam counting from the center. It displays a frequency range of 0 to 50 kHz and showcases two bands. The first band is below 10 kHz, while the second spans approximately from 11 to 47 kHz. Notably, transmission revival is distinctly observable around 25 kHz and 40 kHz, indicating its independence from the spectral bands. Please note that due to resolution limitations, some instances of transmission revival might be challenging to observe without significantly zooming in.

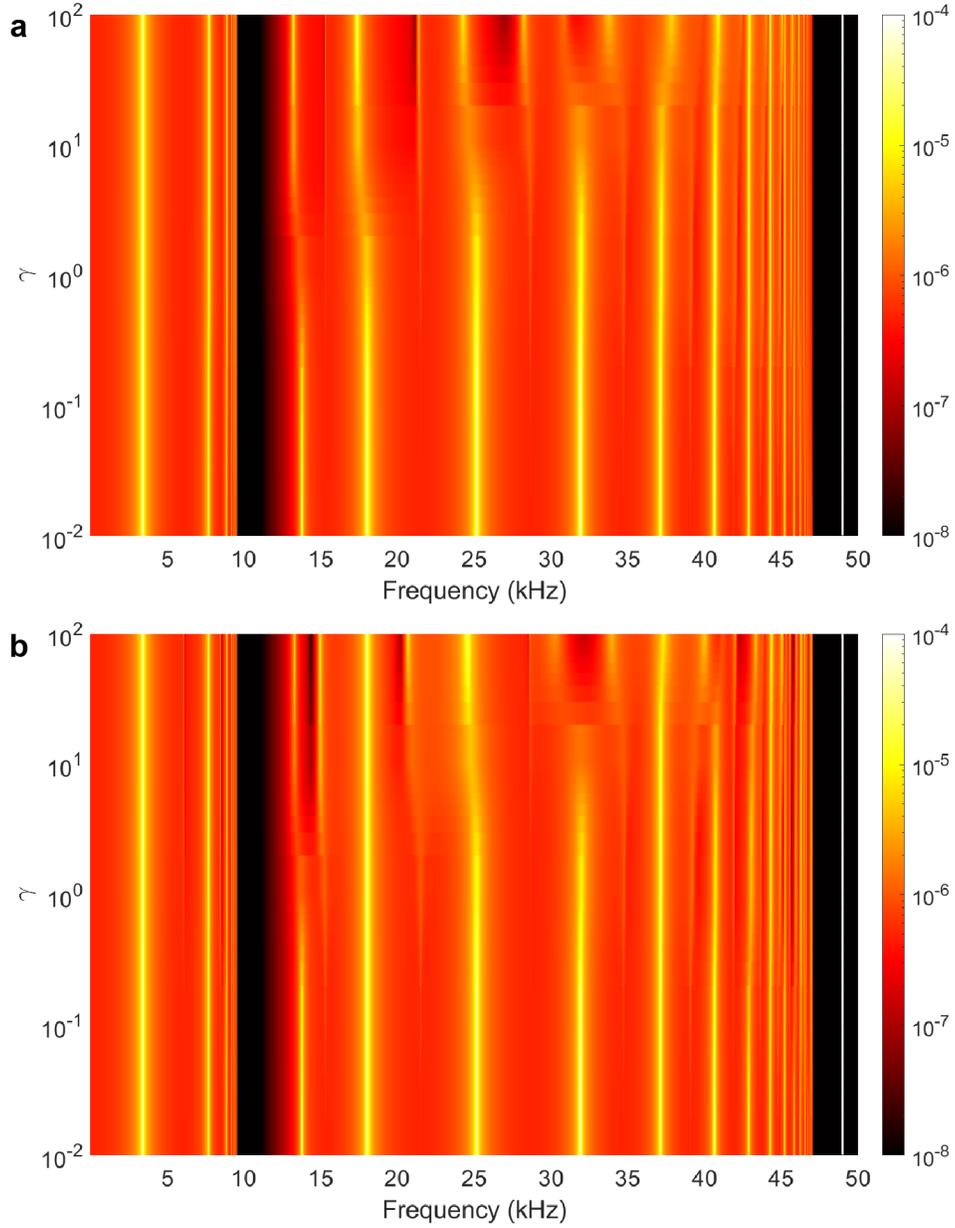

FIG. 6. Density plots of the transmitted signal for an $N_{15}$ metastructure (belongs to the same group of metastructures as shown in Fig. 5) under center excitation when a non-Hermitian defect (where damping intensity $\gamma$ is added) is embedded in different locations. (a) Defect is introduced on its 1st cross beam counting from the center. (b) Defect is introduced on its 5th cross beam counting from the center. The frequency range spans from 0 to 50 kHz and shows two bands, similar to Fig. 5. Similarly, the transmission revival is observable at multiple frequencies, such as 22 kHz, 28 kHz, 35 kHz, etc., emphasizing its independence from system size and defect location. Please note that due to resolution limitations, some instances of transmission revival might be challenging to observe without significantly zooming in.

# APPENDIX B: COUPLED MODE THEORY ANALYSIS

## 1. Introduction

The purpose of this appendix is to detail the coupled mode theory (CMT) analysis to provide a theoretical foundation for the computational and experimental results presented in the main text. The analysis summarized below includes only results obtained using the CMT formulation.

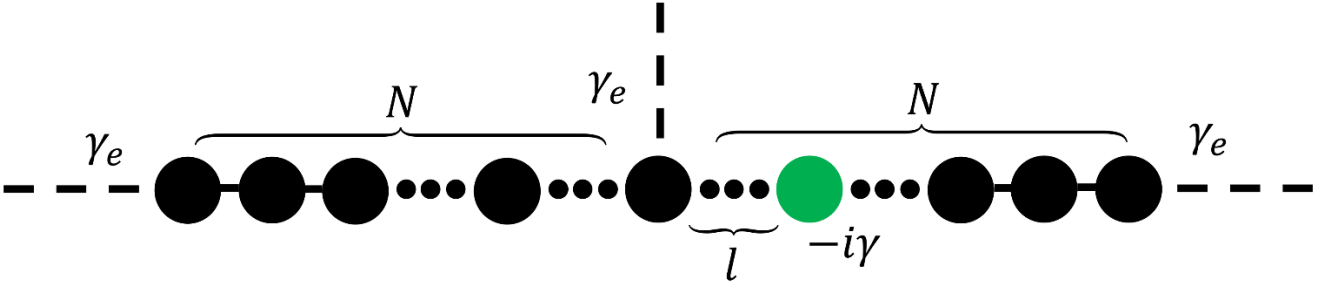

FIG. 7. Tight-Binding Schematic. The system model is described in Eq. (B1) by a chain of $2N + 1$ resonators with uniform coupling $\kappa$ and on-site potentials $\omega_0$, along with loss $\gamma$ introduced to the $N + 1 + l$ site. The isolated system is turned to a scattering set-up by coupling leads to the center or the ends of the array with coupling $\gamma_e$.

## 2. Model Definition and Effective Hamiltonian

The elastodynamics model associated with this research consists of a one-dimensional array of $2N + 1$ individual cross beams arranged transversely on a main connecting beam with a tunable loss element on the $N + l + 1$ cross beam. Considering the fundamental modes associated with the cantilever free vibration associated with the isolated cross beams, the structure can be modeled using CMT as isolated resonances with frequency $\omega_0$ with nearest neighbor coupling $\kappa$ via the main connecting beam. Meanwhile, non-Hermiticity introduced by loss on the $N + l + 1$ element can be incorporated by the imaginary term $-i\gamma$. Therefore, the Hamiltonian operator of the composite closed system is expressed in the basis of the isolated resonances as,

$$\widehat{H} = \sum_{n=1}^{2N+1} \omega_0 |n\rangle\langle n| + \sum_{n=1}^{2N} \kappa |n\rangle\langle n+1| + c.c. - i\gamma |N + l + 1\rangle\langle N + l + 1| \quad (B1)$$

Here $|n\rangle$ represents the isolated mode of the $n^{th}$ cross beam, referred to here on as sites. The system can be turned to a scattering set-up by attaching leads to the respective sites. Fig. 7 shows a tight-binding schematic of the system. The perturbation of the leads can be modeled by an imaginary self-energy correction which results the effective Hamiltonian,

$$\widehat{H}_{eff} = \widehat{H} - \frac{i\widehat{D}\widehat{D}^T}{2} \quad (B2)$$

Here, $\widehat{D} = \sqrt{2\gamma_e}(|1\rangle\langle e_1| + |N+1\rangle\langle e_2| + |2N+1\rangle\langle e_3|)$ is referred to as the coupling matrix, where $|e_m\rangle$ indicates the propagating modes of the $m^{th}$ lead and $\gamma_e$ is the coupling coefficient to the coupled sites inside the structure. Physically, this corresponds to three symmetrically coupled scattering channels at the center and either side of the system. Therefore, the coupled mode theory equations are expressed [26],

$$\omega|\psi\rangle = \widehat{H}_{eff}|\psi\rangle + i\widehat{D}|s^+\rangle \quad (B3a)$$

$$|s^-\rangle = \widehat{D}^T|\psi\rangle - |s^+\rangle \quad (B3b)$$

Here, $\omega$ is the frequency of the interrogating monochromatic signal with incident wavefront $|s^+\rangle = \sum_{m=1}^{3} s_m^+ |e_m\rangle$, $|\psi\rangle = \sum_{n=1}^{2N+1} a_n |n\rangle$ denotes the steady state scattering field inside the structure and $|s^-\rangle = \sum_{m=1}^{3} s_m^- |e_m\rangle$

represents the outgoing wavefront. The first of these equations describes internal dynamics with respect to an inhomogeneous driving term and the second encodes continuity of the input/output waves and scattering field at the interface of the leads. Eq. (B3a) can be re-arranged to find the scattering field resulting from excitation due to the incident wavefront,

$$|\psi\rangle = i\hat{G}\hat{D}|s^+\rangle \tag{B4}$$

where $\hat{G} = (\omega \mathbb{I}_{2N+1} - \hat{H}_{eff})^{-1}$ is the Greens function and $\mathbb{I}_M$ is the $M \times M$ identity matrix. Inserting Eq. (B4) to Eq. (B3b) results the input/output relation,

$$|s^-\rangle = \hat{S}|s^+\rangle \tag{B5}$$

where $\hat{S} = -\mathbb{I}_3 + i\hat{D}^T \hat{G}\hat{D}$ is the scattering matrix.

### 3. Identification of Symmetry Induced Single Sided Resonance

Omitting the explicit calculation, because it is geometrically apparent, the effective Hamiltonian is symmetric with respect to parity inversion in the absence of additional losses ($\gamma = 0$). The parity operator can be expressed in the basis of the sites,

$$\hat{P} = \sum_{n=1}^{2N+1} |n\rangle\langle 2N + 2 - n| \tag{B6}$$

It is well known [29] that parity symmetric operators have eigenvectors that are either symmetric or anti-symmetric. That is, symmetric and anti-symmetric eigenmodes of the effective Hamiltonian, in the absence of additional losses, $\hat{H}_{eff}(\gamma = 0)|\psi_n\rangle = \omega_n|\psi_n\rangle$, represented in the basis of the sites as, $|\psi_n\rangle = \sum_{n=1}^{2N+1} c_n|n\rangle$, satisfy $\hat{P}|\psi_n\rangle = \pm|\psi_n\rangle$, respectively. Consequently, anti-symmetric modes (indicated by $n = n'$) have a nodal point on their center site,

$$\langle N + 1|\psi_{n'}\rangle = -\langle N + 1|\hat{P}|\psi_{n'}\rangle = -\sum_{n=1}^{2N+1} \langle N + 1|n\rangle\langle 2N + 2 - n|\psi_{n'}\rangle = -\langle N + 1|\psi_{n'}\rangle$$

$$\rightarrow c_{N+1} = \langle N + 1|\psi_{n'}\rangle = 0 \tag{B7}$$

The presence of the nodal point in the wavefunctions enforces a single-sided resonance [27,28] because waves incident from the center lead, at the corresponding energy, are blocked from entering the system. To understand this, perform an eigenfunction expansion on the Green's function in Eq. (B4), assuming an incident wave in proximity to an anti-symmetric mode's eigenfrequency, $\omega_{n'}$ at the center lead, $|s^+\rangle = |e_2\rangle$,

$$|\psi(\omega = Re(\omega_{n'}))\rangle = i\sum_{n=1}^{2N+1} \frac{|\psi_n\rangle\langle\psi_n|}{\omega - \omega_n} \hat{D}|e_2\rangle \approx i\sqrt{2\gamma_e}\langle\psi_{n'}|N + 1\rangle|\psi_{n'}\rangle = 0 \tag{B8}$$

The approximate equality assumes lead coupling coefficients are sufficiently small as compared with the level spacing so that $\text{Im}(\omega_n) \approx 0$, so the other terms can be truncated from the summation. Despite $\hat{H}_{eff}$ being a non-Hermitian operator, the above expression uses the appropriately normalized left-eigenvectors for $\langle\psi_n|$, so that a bi-orthogonal inner product can be preserved, $\langle\psi_n|\psi_m\rangle = \delta_{nm}$.

Using Eq. (B3), the validity of the approximation can be tested by computing the occupation values $|a_n|^2$ of the resulting scattering field by projecting it onto the basis of the effective Hamiltonian, $a_n = i\sqrt{2\gamma_e}\langle\psi_n|N + 1\rangle$. The results for excitation at frequencies $\omega = \text{Re}(\omega_n)$ for a symmetric and anti-symmetric mode are displayed in Fig. 8. Notice, the occupation values are many orders smaller for the anti-symmetric mode frequency, as compared with the symmetric mode frequency. Not only this, but the resulting scattering field at the anti-symmetric frequency consists entirely of symmetric eigenmode components.

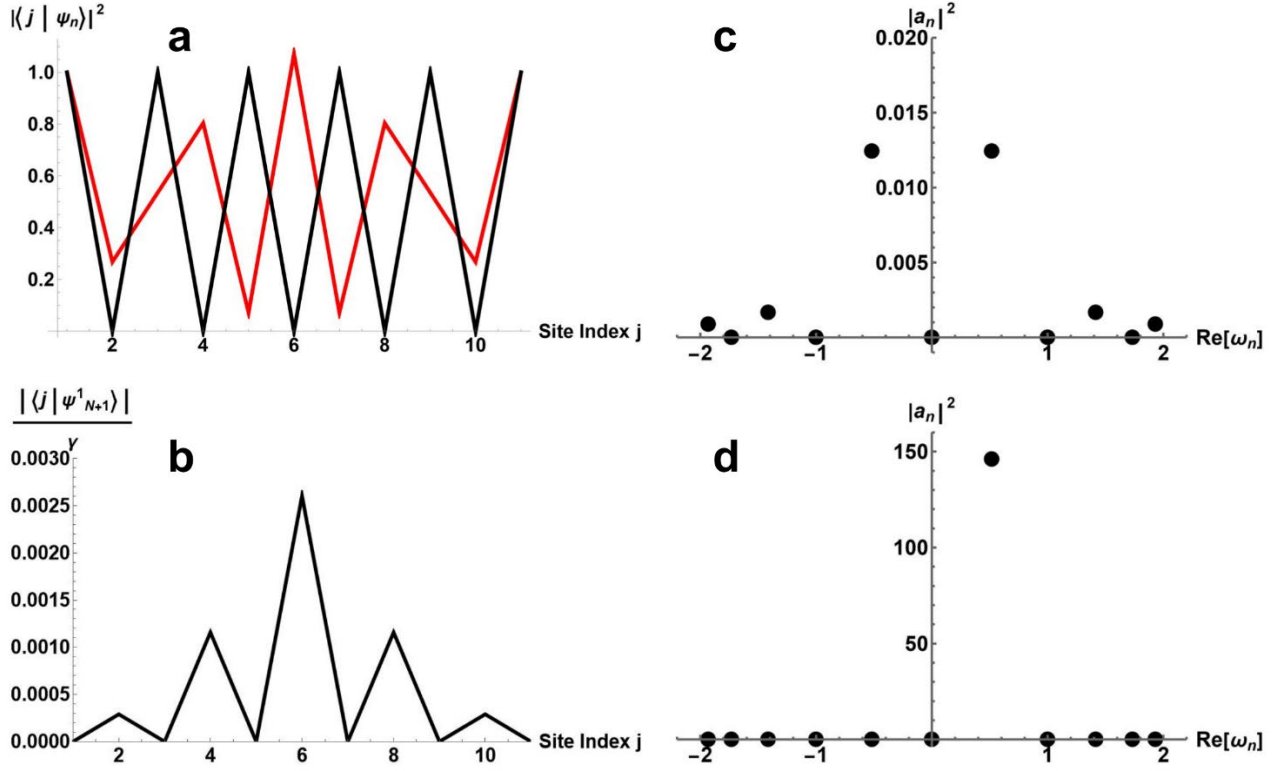

FIG. 8. Eigenmodes of Parity Symmetric Structure and Projection of Center Excitation. (a) Intensity profile of eigenmodes corresponding to the closed Hermitian system with $N = 5$. The black line represents the $n = N + 1$ anti-symmetric mode with a nodal point at the site $j = N + 1$ and the red line corresponds to the $n = N + 2$ symmetric mode which has an anti-node at its center. (b) Norm of first order correction to antisymmetric eigenmode $n = N + 1$ due to non-Hermitian perturbation $-i\gamma$ to the site $j = N + 1 + l$ for $l = 1$, demonstrating its function to destroy the nodal point. (c)(d) Occupation values of the scattering field resulting from center excitation with weak lead coupling $\gamma_e = 10^{-2}$, driven at the real part of the eigenfrequency of the modes (c) $n = N + 1$ and (d) $n = N + 2$ respectively. (b) The anti-symmetric mode frequency indicates the scattering field is composed of only the symmetric modes due to their resonance overlap, while anti-symmetric modes (in particular $n = N + 1$) are excluded from the superposition. (c) The projection at the symmetric mode frequency indicates the excitation is primarily comprised of the symmetric mode nearest to the driving frequency. The excited intensity is many orders smaller in the anti-symmetric case due to the presence of a nodal point at the driven lead.

### 4. First Order Perturbation Theory of Non-Hermitian Element

The eigenvalue problem for the closed Hermitian system $\gamma = 0$ defined by,

$$\omega_n^0 |\psi_n^0\rangle = \hat{H} |\psi_n^0\rangle \quad (B9)$$

has the following analytic solution when $\omega_0 = 0$ and $\kappa = 1$ are set as reference and scale [30],

$$|\psi_n^0\rangle = \frac{1}{\sqrt{2(N+1)}} \sum_{j=1}^{2N+1} \sin\left(\frac{\pi n}{2(N+1)} j\right) |j\rangle; \quad \omega_n^0 = 2\cos\left(\frac{\pi n}{2(N+1)}\right) \quad (B10)$$

Specifically, wavefunctions possessing nodal points in the center of their profile occur when $n$ is even,

$$\langle N+1|\psi_n^0\rangle = \frac{1}{\sqrt{2(N+1)}}\sum_{j=1}^{2N+1} \sin\left(\frac{\pi n}{2(N+1)}j\right)\langle N+1|j\rangle = \frac{1}{\sqrt{2(N+1)}}\sin\left(\frac{\pi n}{2}\right) \quad (B11)$$

To be more explicit, we will focus on the mode $n = N+1$ where $N$ will be assumed odd unless otherwise specified, for which,

$$|\psi_{N+1}^0\rangle = \frac{1}{\sqrt{2(N+1)}}\sum_{j=1}^{2N+1} \sin\left(\frac{\pi}{2}j\right)|j\rangle; \quad \omega_{N+1}^0 = 2\cos\left(\frac{\pi}{2}\right) = 0 \quad (B12)$$

First order non-degenerate perturbation theory prescribes the first order correction to the wavefunction [31],

$$|\psi_{N+1}^1\rangle = \sum_{\substack{n=1 \\ n\neq N+1}}^{2N+1} \frac{\langle \psi_n^0|\hat{V}|\psi_{N+1}^0\rangle}{\omega_{N+1}^0 - \omega_n^0}|\psi_n^0\rangle \quad (B13)$$

The perturbation matrix is $\hat{V} = -i\gamma|N+1+l\rangle\langle N+1+l|$ and the first order correction can be calculated as,

$$|\psi_{N+1}^1\rangle = i\gamma \sum_{\substack{n=1 \\ n\neq N+1}}^{2N+1} \frac{\langle \psi_n^0|N+1+l\rangle\langle N+1+l|\psi_{N+1}^0\rangle}{\omega_n^0}|\psi_n^0\rangle$$

$$= \frac{i\gamma}{(2(N+1))^{\frac{3}{2}}}\sum_{\substack{j,n=1 \\ n\neq N+1}}^{2N+1} \frac{\sin\left(\frac{\pi n}{2(N+1)}(N+1+l)\right)\sin\left(\frac{\pi}{2}(N+1+l)\right)}{\omega_n^0}\sin\left(\frac{\pi n}{2(N+1)}j\right)|j\rangle \quad (B14)$$

which is non-zero when $l$ is odd, where the unperturbed anti-symmetric eigenfunction acquires its anti-nodes. Assuming this to be the case,

$$|\psi_{N+1}^1\rangle = \frac{i\gamma\,(-1)^{\frac{N+l}{2}}}{(2(N+1))^{\frac{3}{2}}}\sum_{\substack{j,n=1 \\ n\neq N+1}}^{2N+1} \frac{\sin\left(\frac{\pi n}{2}+\frac{\pi n l}{2(N+1)}\right)}{\cos\left(\frac{\pi n}{2(N+1)}\right)}\sin\left(\frac{\pi n}{2(N+1)}j\right)|j\rangle \quad (B15)$$

We see the partial effect of the non-Hermitian perturbation is to introduce a phase change to the eigenmodes of the system which results in an interference between the modes of the scattering field. The norm of the correction is plotted in Fig. 9, where we see a strong disruption of the nodal point. Of course, for weak coupling to the transmission lines, the above results pertains also to the shape of resonant modes.

Inspecting the effect of the perturbation on the nodal point, we find,

$$\langle N+1|\psi_{N+1}^1\rangle = \frac{i\gamma\,(-1)^{\frac{N+l}{2}}}{2(2(N+1))^{\frac{3}{2}}}\sum_{\substack{n=1 \\ n\neq N+1}}^{2N+1} \frac{\sin\left(\frac{\pi n}{2}+\frac{\pi n l}{2(N+1)}\right)}{\cos\left(\frac{\pi n}{2(N+1)}\right)}\sin\left(\frac{\pi n}{2}\right) \quad (B16)$$

Despite being unable to evaluate the summation in a closed form, a numerical investigation for various odd values of $N$ with respect to odd values of $l$ has indicated the nodal point destruction is independent of the defect placement and the size of the system, up to the normalization of the wavefunction due to the increasing number of sites.

Meanwhile, the first order perturbation to the eigenvalues is provided by,

$$\omega_n^1 = \langle \psi_n^0|\hat{V}|\psi_n^0\rangle \quad (B17)$$

which evaluates to,

$$\omega_n^1 = -i\gamma|\langle N+1+l|\psi_n^0\rangle|^2 = -\frac{i\gamma}{2(N+1)}\sin^2\left(\frac{\pi n}{2}+\frac{\pi l}{2(N+1)}\right) \quad (B18)$$

Therefore, to first order, a non-Hermitian perturbation only has the effect of increasing the linewidth of the modes, without changing their resonant frequency. Meanwhile, a Hermitian perturbation would introduce a resonant shift to the mode frequencies, which is more inconvenient to track experimentally.

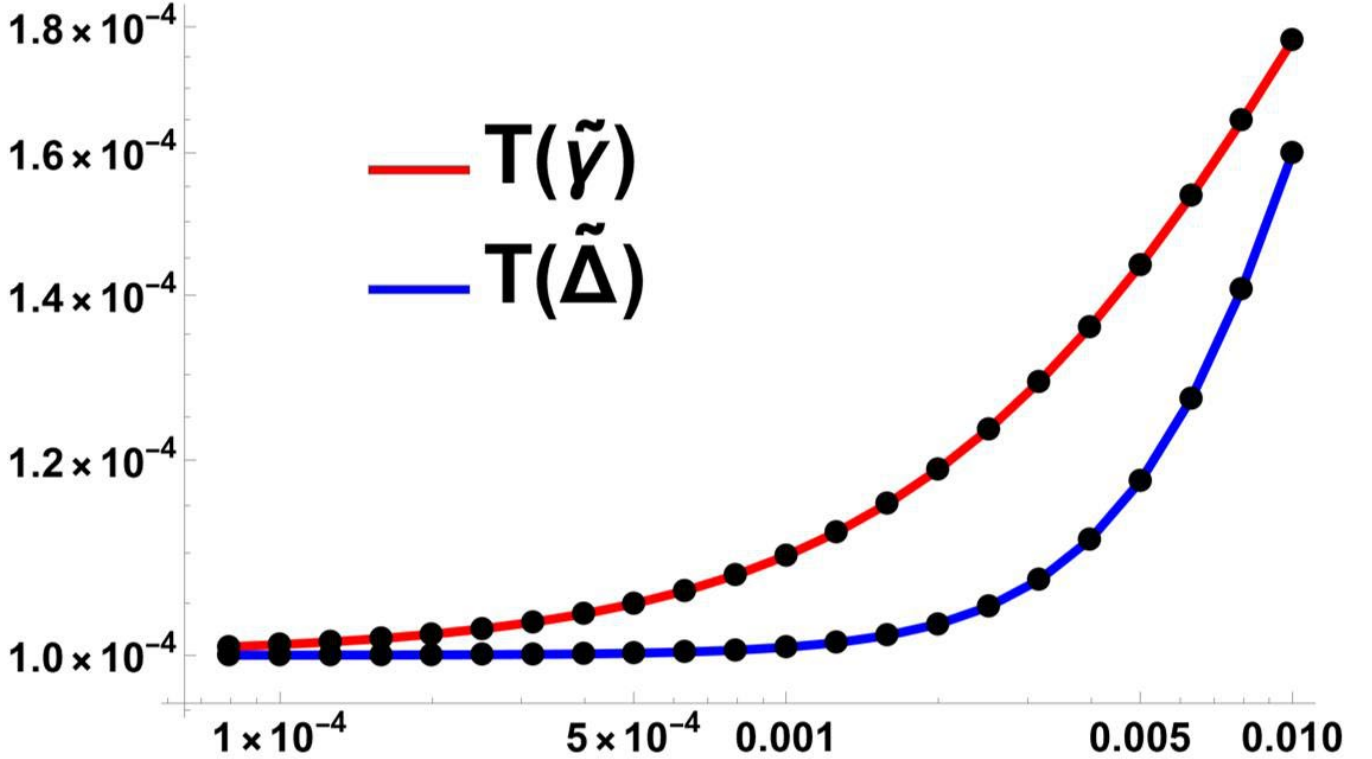

FIG. 9. Transmission Resurgence for Lossy and Hermitian Defects. The vertical axis is transmission from the center lead to the leftmost lead and the horizontal axis indicates the perturbation strength of the defect. The red line is Eq. (B21b) plotted with respect to the strength of the non-Hermitian defect, $\tilde{\gamma}$. The blue line is Eq. (B22b) plotted with respect to the strength of the Hermitian defect, $\tilde{\Delta}$. The black points are numerical results simulated with weak coupling $\tilde{\gamma}_e = 0.01$, for a system size $2N + 1$ with $N = 5$, and defect placement $l$ to the right of the center site with $l = 1$ to verify the analytic expressions, however the expressions are independent of odd values of $N$ and $l$. Notably, the transmission enhancement is linear with respect to $\tilde{\gamma}$ and quadratic with respect to $\tilde{\Delta}$. Therefore, a non-Hermitian perturbation to the defect, counterintuitively, results in a more sudden increase in transmission than that of a Hermitian perturbation. This underscores the utility of simultaneously violating both geometric and hidden (time-reversal) symmetries in the design and application of wave mechanical devices.

### 5. Analytic Greens Function and Transmission Revival

In this section, we calculate the transmission to the left-most lead from an incident signal at the center lead. Specifically, we are interested in the quantity $T = |\langle e_1|s^-\rangle|^2$ due to the single port excitation defined by $|s^+\rangle = |e_2\rangle$, driven at the eigenfrequency of an anti-symmetric eigenmode of the unperturbed system (specifically $\omega = \omega_0 = 0$ with $N$ and $l$ both odd). At this frequency, the Greens function reduces to $\hat{G} = -(H_{eff})^{-1}$, corresponding to a tri-diagonal matrix inversion. Since the coupling is assumed to be uniform, matrix elements of the inverse can be evaluated analytically using [32],

$$-(H_{eff})^{-1}_{ij} = \begin{cases} (-1)^{i+j}\kappa^{|j-i|}\theta_{i-1}\phi_{j+1}/\theta_{2N+1}, & i<j \\ \theta_{i-1}\phi_{j+1}/\theta_{2N+1}, & i=j \\ (-1)^{i+j}\kappa^{|i-j|}\theta_{j-1}\phi_{i+1}/\theta_{2N+1}, & i<j \end{cases} \quad (B19)$$

Along with the following set of recursion relations,

$$\theta_i = d_i \theta_{i-1} - \kappa^2 \theta_{i-2}; \quad i = 2, 3, \ldots, 2N+1 \tag{B20a}$$

$$\phi_i = d_i \phi_{i+1} - \kappa^2 \phi_{i+2}; \quad i = 2N, 2N-1, \ldots, 1 \tag{B20b}$$

where $d_i = -i[\gamma_e(\delta_{i,1} + \delta_{i,N+1} + \delta_{i,2N+1}) + \gamma \delta_{i,N+1+l}]$ are the diagonal elements of $\hat{H}_{eff}$ and the initial terms of the sequence are provided by $\theta_0 = 1$, $\theta_1 = -(H_{eff})_{11}$ and $\phi_{2N+2} = 1$, $\phi_{2N+1} = -(H_{eff})_{2N+1,2N+1}$. Once the Greens function is obtained, an expression for the transmission coefficient can be obtained,

$$t = 2i\tilde{\gamma}_e \frac{\tilde{\gamma}_e + \tilde{\gamma}}{(\tilde{\gamma}_e + \tilde{\gamma})(\tilde{\gamma}_e^2 + 1) + \tilde{\gamma}_e} \tag{B21a}$$

where the tilde here indicates the quantity is normalized by the coupling between sites of the system (i.e., $\tilde{\gamma} \equiv \gamma/\kappa$ and $\tilde{\gamma}_e \equiv \gamma_e/\kappa$). Correspondingly, the transmission can be computed using Eq. (B5). The result is,

$$T = \left[\frac{2\tilde{\gamma}_e(\tilde{\gamma}_e + \tilde{\gamma})}{(\tilde{\gamma}_e + \tilde{\gamma})(\tilde{\gamma}_e^2 + 1) + \tilde{\gamma}_e}\right]^2 \tag{B21b}$$

As a monotonically increasing function in $\tilde{\gamma}$, Eq. (B21b) demonstrates enhancement in the transmission as a response to increasing strength of the non-Hermitian defect. Notice this result is independent of the size of the system and the defect placement, provided both $N$ and $l$ are odd. The reason the transmission does not vanish for $\gamma = 0$, despite $\hat{H}_{eff}$ being parity symmetric, is due to overlapping resonances which are broadened by increasing the lead coupling, which is particularly significant for large values of $\tilde{\gamma}_e$. This fact regarding overlapping resonances can also be recognized from the occupation values presented in Fig. 8.

### 6. Impact of Hidden Symmetry Violation

On the surface, the non-Hermitian perturbation only serves to violate the parity symmetry of the underlying structure. In fact, this is not the case. As discussed in the above perturbation theory analysis, additional loss mixes the phases associated with the unperturbed eigenmodes. Moreover, in the expression for the transmission coefficient, Eq. (B21a), the effect of a Hermitian defect can be studied by substituting the added loss with a resonant shift of the isolated mode $\Delta$, i.e., $-i\tilde{\gamma} \to \tilde{\Delta}$. In this case, the transmission coefficient is,

$$t = 2i\tilde{\gamma}_e \frac{\tilde{\gamma}_e + i\tilde{\Delta}}{(\tilde{\gamma}_e + i\tilde{\Delta})(\tilde{\gamma}_e^2 + 1) + \tilde{\gamma}_e} \tag{B22a}$$

So, from Eq. (B5), the transmission is computed as,

$$T = \frac{4\tilde{\gamma}_e^2(\tilde{\gamma}_e^2 + \tilde{\Delta}^2)}{(\tilde{\gamma}_e^2 + \tilde{\Delta}^2)(\tilde{\gamma}_e^2 + 1)^2 + 2\tilde{\gamma}_e^2(\tilde{\gamma}_e^2 + 1) + \tilde{\gamma}_e^2} \tag{B22b}$$

The key feature to observe is, while Eq. (B21b) indicated an enhancement in the transmission that is first order in $\tilde{\gamma}$, the transmission enhancement from Eq. (B22b) is only second order in $\tilde{\Delta}$, due to the vanishing cross-term. Therefore, in addition to violating parity symmetry, the non-Hermitian perturbation has the profound effect of also violating the hidden time-reversal symmetry which results in a considerable impact on the transmission revival, as demonstrated in Fig. 9. Furthermore, in the limit of extremely large defect strengths both expressions approach the same value for the transmittance. This is not surprising because, in either case, the second half of the system becomes decoupled entirely. Such a scenario is unrealistic in the experimental platform, where we consider small perturbations, i.e., $\tilde{\gamma}, \tilde{\Delta} \ll 1$.